\documentstyle[twoside,fleqn,espcrc2]{article}

\def\be{\begin{equation}}
\def\ee{\end{equation}}
\def\al{\alpha}

\newcommand{\AmS}{{\protect\the\textfont2
  A\kern-.1667em\lower.5ex\hbox{M}\kern-.125emS}}

\title{ Lipatov Pomeron and the Heisenberg chain }

\author{
        R. Janik  and J. Wosiek\address{Institute of Physics, 
         Jagellonian University , \\
        Reymonta 4, Cracow, Poland}
        \thanks{Supported in part by the Stefan Batory Foundation and the 
            KBN grants no 
                    PB 2P03B19609 and PB 2P30225206.}
        \thanks{Presented by J. Wosiek. }  }
\begin{document}

\begin{abstract}
It is pointed out that the recently discovered analogy between the
QCD description of the Pomeron exchange and the Heisenberg chain
may allow applying lattice techniques to the high energy phenomena.   
\end{abstract}

\maketitle

\section{HIGH ENERGY SCATTERING IN THE REGGE LIMIT}
Customary lattice methods are considered suitable for
the description of the static, or low energy, nonperturbative 
properties of hadrons. Even though some progress has been made
in studying the scattering length, form factors and the deep inelastic
structure functions, the finite ultraviolet cut-off $a^{-1}\sim 2 GeV$ 
limits the application of classical lattice techniques 
mainly to the static phenomena. However, there exists an intriguing
mapping, between the description of a high energy scattering and the
one dimensional Heisenberg chain, which offers a possibility to use the
methods of statistical physics to study the genuine high energy scattering.  

The typical high energy scattering is described by the
Regge amplitude
\be
A(s,t)\sim s^{\alpha(t)},\;\;\;\;s >> -t,
\ee
where $s$ and $t$ are the Mandelstam variables and $\alpha(t)$ describes
the Regge trajectory exchanged in the
$t$ channel. Phenomenologically Regge trajectories are linear
\be
\alpha(t)=\alpha_0+\alpha' t,
\ee
with the intercept, $\al_0$, and the slope, $\al'$, depending on the
quantum numbers exchanged in the $t$ channel.
The total cross section
\begin{equation}
\sigma_{tot}={1\over s} Im A_{el}(s,0),
\end{equation} 
is given in terms of the elastic scattering in the forward direction. 
The latter is described by the exchange of the Pomeranchuk trajectory
with the vacuum quantum numbers. Unitarity restricts the
growth of the cross sections. The Froissart bound
\begin{eqnarray}
\sigma_{tot} \leq const. \log^2{(s)},
\end{eqnarray}
limits the intercept $\al_0 \leq 1$. 
 Regge behaviour occurs also in the deep inelastic scattering
 at small $x$
\begin{eqnarray}
F(x,Q^2)={1\over s}Im A_{\gamma^{\star}P}(s,t=0) && \\
s={Q^2\over x},\;\;\; x={Q^2\over 2M\nu}, \;\;\; x \ll 1 .  & &
\end{eqnarray}
Again unitarity limits the growth of the structure functions
at small $x$
\be
F(x,Q^2) \leq const. \log^2{(x)}.
\ee
Emergence of the Regge behaviour is qualitatively  well understood  
as the result of the exchange of the ladder diagrams \cite{pol}.
 The quantitative derivation of the Pomeron exchange in QCD
 was pioneered by Lipatov and is presently actively
investigated \cite{LipRev,FadKor94,Kor95}. 

\section{POMERON IN QCD }

Working in the leading logarithmic approximation, $\al_s\ll 1,\;\;\;
\al_s \log{(s/M^2)} \sim 1$, Lipatov and others \cite{KLFBL} 
have identified and summed the
relevant class of the QCD ladder diagrams in the Regge limit. In addition to
the simple ladder diagrams, other diagrams also contribute and the whole
class is commonly referred to as the exchange of two reggezied gluons.
The result \cite{KLFBL} reads
\begin{eqnarray}
A_{LLA}\sim s^{1+\Delta_{BFKL}(t)},&& \\ 
\Delta_{BFKL}(0)={4 \al_s N_c\over \pi}\log{(2)}  && \label{al}.
\end{eqnarray}
Therefore QCD reveals the Regge behaviour of the elastic amplitude
and the intercept of the Pomeranchuk trajectory is known. However
the amplitude (\ref{al}) violates unitarity, hence a lot of subsequent work
has been devoted to find unitarity corrections to Eq.(\ref{al}). In particular
the scheme termed generalized leading logarithmic approximation (GLLA) 
was proposed \cite{Bar80,CheDick81} 
which identifies the non-leading contributions which should restore 
unitarity. In the GLLA the Mellin transform $\tilde{A}(\omega,t)$ of
 the scattering amplitude
\be
A(s,t) = is\int_{\delta-i\infty}^{\delta+i\infty} {d\omega\over 2\pi i}
\left({s\over M^2}\right)^{\omega} \tilde{A}(\omega,t), \label{mel}
\ee
is given as a sum of contributions $\tilde{A}_n$ from the exchange of 
 $n$ reggeons (reggeized gluons) in the $t$ channel. For fixed $n$
they are given by the evolution operator $T_n(\omega)$
\begin{eqnarray*}
\lefteqn{ \tilde{A}_n(\omega,t)
=\int d^2k_1\dots d^2k_n 
d^2k_{1}'\dots d^2k_{n}' 
}
\\
&& \Phi_A(\{k\};q)
T_n(\{k\},\{k'\});\omega) \Phi_B(\{k'\};q).
\end{eqnarray*}
Where $\Phi_{A(B)}$ are the wave functions of the hadron $A(B)$ in the
reggeon basis. The evolution operator conserves the number of reggeons
and satisfies the following Lippmann-Schwinger equation
\be
\omega T_n(\omega)=T_n^0(\omega) + {\cal{H}}_n T_n(\omega),
\ee
which can be solved symbolically as
\be
T_n(\omega)={1\over \omega-{\cal{H}}_n } T_n^0(\omega). \label{sol}
\ee
Comparing Eqs.(\ref{sol}) and (\ref{mel}) we see that the largest eigenvalue
of the Hamiltonian of $n$-reggeons determines the high energy behaviour
of the scattering amplitude. Moreover in the large $N_c$ limit only planar 
diagrams give leading contribution, and consequently ${\cal{H}}_n$ describes
the system of $n$ degrees of freedom which possesses only the nearest neighbour 
interactions. At that point the analogy with statistical systems emerges.
It becomes even more appealing after transforming to the impact parameter
representation, $\vec{k_i} \rightarrow \vec{b}_i$, and after replacing transverse
variables by the complex coordinates, $\vec{b}_i=(x_i,y_i)\rightarrow
z_i=x_i+i y_i, \overline{z}_i=x_i-i y_i$. It was shown 
\cite{Lip93JETP,FadKor94} that the 
resulting Hamiltonian describes the linear chain of the quantum spins
with the nearest-neighbours interactions in a complete analogy to the
Heisenberg chain of ordinary spins. 
  There are two complications, however,
which make this problem nontrivial and challenging. First, the spin operators 
in question are infinitely dimensional. In fact they generate the $s=0$ 
unitary representation of the $SL(2,C)$ group. Second, the nearest neighbours
interaction is nonlocal in the $z,\overline{z}$ space. In spite of this 
complications, Faddeev and Korchemsky \cite{FadKor94}, using the 
generalized Bethe ansatz  have confirmed
Lipatov's hypothesis \cite{Lip90}, 
that the n-reggeon problem is holomorphically separable and 
 in principle, exactly solvable. The $n=2$ hamiltonian was diagonalized 
explicitly
and BFKL slope, Eq.(\ref{al}) was reproduced.
    
\section{THE ODERON CASE}

 The $n=3$ case, which corresponds to phenomenologically important 
oderon problem, is solved only partially. Analytical results are available
 for integer values $(h=n)$ of the Casimir operator of two spins,
$\hat{q}_2=(s_i+s_j)^2=-h(h-1)$. Also the asymptotic expansion of the
maximal eigenvalue of ${\cal{H}}_3$ in $1/h$ was derived. All these results
are based on the Bethe ansatz approach. We have studied the solution $Q_3(z)$
of the corresponding Yang-Baxter equation by analogy to the standard theory
of Fuchsian equations of the second order. Even though the Yang-Baxter equation
for $Q_3(z)$ is of the third order, the elements of the general theory apply.
In particular, we have found that the case considered corresponds to the
degenerate situation with the characteristic exponent $r=0$. Hence, the equation 
has one regular and two irregular solutions around $z=0$. This is independent
of the eigenvalues of the two Casimir operators $\hat{q}_2$ and $\hat{q}_3$ 
which parametrize the problem. The coefficients $f_n$ of the power expansion
\be
Q_3(z)=\sum_n f_n z^n,  \label{pow}
\ee
are determined by the following recursion relation
\begin{eqnarray}
(n+1)^3 f_{n+1}= &&  \nonumber \\ 
 \left(      n((2n+1)(n+1)+q_2)+iq_3    \right) f_n && \nonumber \\
-(n-1)(n(n+1)+q_2)f_{n-1}.&&  \label{rec}
\end{eqnarray}
This recursion is equivalent to the recursion derived in Ref.\cite{FadKor94}
where the expansion in terms of the Legendre polynomials was studied.
In particular, for integer $h$ $Q_3(z)$ is a finite polynomial in $z$
only for the discrete values of $q_3$.
The quantization condition of $q_3$ which follows from Eq.(\ref{rec}) 
is the same as one derived in Ref.\cite{Kor95}
 The straightforward power expansion (\ref{pow}) may, however,
prove more suitable for the study of the analytic structure of the solution
and in consequence may allow the analytic continuation for arbitrary~$h$.   

\section{LATTICE FORMULATION}

In closing we would like to comment on the feasibility of applying
lattice Monte Carlo methods to solve this problem numerically.

 Contrary to the standard applications, physically interesting
questions arise already for the {\em finite} number of exchanged reggeons, i.e. for
finite length of the Heisenberg chain, say $n=3$. However the volume of
the transverse parameter space is infinite, hence the normal difficulties
associated with this infinite volume limit would arise. In fact 
the hamiltonian ${\cal{H}}_n$ has the global conformal invariance, therefore
 the critical slowing down is expected 
to occur. So, the challenge is how to simulate
effectively a conformal theory on the lattice?  

 The nearest neighbours interaction is nonlocal which also complicates
 practical applications. Since, however, the system is critical, one is bound 
to employ the nonlocal (cluster) algorithms. Therefore the nonlocality
of ${\cal{H}}_n$ may not be the main problem.

 The choice of the representation may help in designing  a 
practical approach.
The "$z$" representation is the one possibility. However one may try to employ
the $SL(2,C)$ symmetry to simplify the problem. Note that one of the analytical
solutions of the Pomeron $n=2$ case uses solely the group theoretical 
information on the spectrum of the unitary representations of the $SL(2,C)$ 
group. 

 Finally, even though the problem is certainly
far from trivial, we should be aware that the $n=2$ case was solved 
analytically, hence the spectrum of the excitations of elementary interaction
is known. Combining this knowledge with the recent progress in 
simulating critical
systems may lead to a realistic proposal for the numerical approach.

\vspace*{.75 cm}
J. W. thanks G. Korchemsky, L. Lipatov and M. A. Nowak for the discussion.

\end{document}